\documentclass{mn2e}
\usepackage{epsf}

\def\msun{{\rm M_{\odot}}}

\title[Growing Supermassive Black Holes by Chaotic Accretion]
{Growing Supermassive Black Holes by Chaotic Accretion}

\author[A. R. King and J. E. Pringle]
{A. R. King$^1$ and
J.E. Pringle$^{1,2}$\\ $^1$Theoretical Astrophysics Group, University
of Leicester, Leicester LE1 7RH\\ $^2$Institute of Astronomy,
University of Cambridge, Madingley Road, Cambridge CB3 0HA}

\date{\today}

\volume{000}

\pagerange{\pageref{firstpage}--\pageref{lastpage}} \pubyear{2006}

\begin{document}

\label{firstpage}

\maketitle

\begin{abstract}

We consider the problem of growing the largest supermassive black
holes from stellar--mass seeds at high redshift. Rapid growth without
violating the Eddington limit requires that most mass is gained while
the hole has a low spin and thus a low radiative accretion
efficiency. If, as was formerly thought, the black--hole spin aligns
very rapidly with the accretion flow, even a randomly--oriented
sequence of accretion events would all spin up the hole and prevent
rapid mass growth. However, using a recent result that the
Bardeen--Petterson effect causes {\it counter}alignment of hole and
disc spins under certain conditions, we show that holes can grow
rapidly in mass if they acquire most of it in a sequence of randomly
oriented accretion episodes whose angular momenta $J_d$ are no larger
than the hole's angular momentum $J_h$. Ultimately the hole has total
angular momentum comparable with the last accretion episode. This
points to a picture in which the accretion is chaotic on a lengthscale
of order the disc size, that is $\la 0.1$ pc.

\end{abstract}


\section{Introduction}

There is recent observational evidence (Barth et al., 2003, Willott et
al, 2003) for supermassive black holes (SMBH) with masses $M \ga
5\times 10^9\msun$ at redshift $z \simeq 6$ . The existence of such
large masses only $\sim 10^9$~yr after the Big Bang is a challenge to
theory. For if we accept that SMBH are largely made of matter which
underwent luminous accretion (Soltan, 1982; Yu \& Tremaine, 2002), and
that the accretion luminosity cannot exceed the Eddington value
\begin{equation}
L_{\rm Edd} = {4\pi GMc\over \kappa} \simeq 10^{47}M_9\, {\rm erg\, s}^{-1}
\end{equation}
(where $M=10^9M_9\msun$ and we have taken $\kappa$ to be the electron
scattering opacity) there is a limit to the rate at which a black hole
can accrete. This comes from setting
\begin{equation}
L_{\rm Edd} = \epsilon c^2\dot M_{\rm acc}
\end{equation}
where $\epsilon$ is the accretion efficiency, specified by the
fractional binding energy of the innermost stable circular orbit
(ISCO) about the hole. The black hole mass then grows as mass--energy
is accreted, at the rate
\begin{equation}
\dot M = (1 - \epsilon)\dot M_{\rm acc}.
\end{equation}
Thus
\begin{equation}
\dot M = {1 - \epsilon\over \epsilon}{M\over t_{\rm Edd}}   
\label{mdot}
\end{equation}
with
\begin{equation}
t_{\rm Edd} = {\kappa c\over 4\pi G} = 4.5\times 10^8~{\rm yr}.
\end{equation}

In general $\epsilon$ varies as accretion proceeds (see below), but
its minimum value $\epsilon_{\rm min}$ over the hole's accretion
history sets a limit to the black hole mass $M$ which can grow from an
initial seed value $M_0$ in a given time. Integrating eqn(\ref{mdot})
we have
\begin{equation}
{M\over M_0} < \exp\biggl[{1\over \epsilon_{min}} - 1\biggr] \,
    \left({t\over t_{\rm Edd}}\right).
\end{equation}
At redshift $z \simeq 6$ the last factor in the exponent is $\simeq
2$. We see that large values of $\epsilon_{\rm min}$ severely restrict
the growth of SMBH. Thus with $\epsilon_{\rm min} =0.43$, as
appropriate for a maximally rotating hole with dimensionless Kerr
parameter $a\simeq 1$ (see below) we find $M/M_0
\la 20$. In this case there is clearly no prospect of growing the
inferred SMBH masses $\sim 5\times 10^9\msun$ at redshift $z \simeq 6$
from stellar--mass seed holes, and one would have to consider other
possibilities (e.g. Volonteri \& Rees, 2005; Begelman et al.,
2006). Growth from seed holes of mass $10\msun$ requires
$\epsilon_{\rm min} \la 0.08$, corresponding to rather lower
black--hole spin rates $a \la 0.5$. Still lower values of $a$ are
desirable if we wish to avoid the difficulty that accretion must be
almost continuous to build up the observed mass.

\section{Keeping the spin low}

In the last Section we showed that growing very large SMBH masses at
high redshift from stellar--mass seeds requires {\it low accretion
efficiency}, or equivalently, {\it modest black--hole spins}. This
runs directly counter to the usual expectation (e.g. Volonteri et al.,
2005; Madau, 2004) that gas accretion produces systematic
spin--up. Volonteri et al. (2005) estimated from semi--analytic
cosmological modelling that the fractional change of mass during each
accretion episode of a growing black hole is quite large, i.e. $\Delta M/M
\sim 1 - 3$. The work of Scheuer \& Feiler (1996) and of 
Natarajan \& Pringle (1998) suggested that for accretion via a
geometrically thin disc the combination of the Lense--Thirring effect
with viscous dissipation (Bardeen \& Petterson, 1975) would always
align the SMBH spin with the angular momentum of the accreting gas on
a timescale typically much shorter than the accretion timescale for
mass and angular momentum. Thus every accretion episode would rapidly
become a spin--up episode.

Recently however King et al. (2005) (hereafter KLOP) showed on quite
general analytic grounds that the Lense--Thirring effect instead
produces {\it counter}alignment on similarly short timescales in
particular cases, namely those where the magnitudes of the angular
momenta $J_h, J_d$ of the hole and disc obey
\begin{equation}
\theta > \pi/2,\ \  J_d < 2J_h. 
\label{cond}
\end{equation}
with $\theta$ is the angle between the vectors ${\bf J}_h$ and ${\bf
J}_d$. Thus if $J_d < 2J_h$, then for ${\bf J}_h, {\bf J}_d$ in random
directions, counteralignment occurs in a fraction
\begin{equation}
f_c = {1\over 2}\biggl[1 - {J_d\over 2J_h}\biggr]
\end{equation}
of cases. We note that counteralignment predominantly involves a shift
in the angular momentum of the hole to counteralign with that of the
disc (see also Lodato \& Pringle, 2006).

It follows that if most gas accretion occurs in randomly oriented
episodes with $J_d \la J_h$, spinup and spindown episodes tend to
alternate in some random way. If instead accretion episodes generally
have $J_d \ga J_h$ the hole must consistently spin up, even if hole
and disc are initially counteraligned (see Lodato \& Pringle 2006 for
details). Clearly the first type of accretion, i.e. with $J_d
\la J_h$, offers the better chance of keeping the black hole spin and
accretion efficiency low, and thus enabling the building large black
hole masses in a short time. In fact for similar values of accreted
mass, the spindown effect of counteraligned accretion is significantly
more effective that the spinup from aligned accretion. This is a
straightforward consequence of the fact that the ISCO for retrograde
rotation is always larger than that for aligned rotation.  Writing the
Boyer--Lindquist coordinate radius of the ISCO as $r = xc^2/GM$, we
have $x = 9, 6$ and $1$ for dimensionless Kerr parameters $a = -1, 0$
and $+1$. The results of Bardeen (1970) show that
\begin{equation}
\epsilon = 1 - \biggl[1 - {2\over 3x}\biggr]^{1/2}
\end{equation}
and that the ISCO has specific angular momentum
\begin{equation}
j = {2\over 3\surd{3}}{GM\over c}[1 + 2(3x-2)^{1/2}].
\end{equation}
We thus see that that for $a = -1, 0, 1$, the specific angular momentum $j$ 
is in the ratio $11:9:3$, illustrating the point above that spindown is
considerably more efficient than spinup.

The general connection between $a$ and $x$ is
\begin{equation}
a = {x^{1/2}\over 3}[4 - (3x-2)^{1/2}].
\end{equation}
Given a hole specified by initial mass and spin parameters $(M_1, x_1)$,
Bardeen (1970) shows that accretion from the ISCO causes these parameters to
evolve as 
\begin{equation}
{x\over x_1} = \biggl({M_1\over M}\biggr)^2.
\end{equation}

Thus to discover how the mass and spin of an SMBH evolve over a series
of accretion episodes, we have to connect the corresponding sequence
of $(M, x)$ values, bearing in mind that the spin parameter $a$
changes sign (hence changing $x$ discontinuously) if accretion
switches between prograde and retrograde states.

\section{The accretion episodes}

Volonteri et al. (2005; see also Wilson \& Colbert, 1995; Hughes \&
Blandford, 2003) showed that if successive accretion episodes add
angular momentum to the growing hole at random angles, then the
eventual spin of the hole is small. We have argued that a small spin
is beneficial in helping the hole to grow more rapidly. But moreover
we argue that if hole growth occurs through a series of accretion
episodes with $J_d \la J_h$, then the spindown process is more
efficient than envisaged by Volonteri et al. (2005). In that paper the
black--hole spin was assumed to align very rapidly with the accretion
flow. Then significant spindown would require a matter supply whose
angular momentum reversed on timescales short compared with the
timescale $t_{\rm al}$ for alignment/counteralignment, which is only a
few times $10^4$~yr (Natarajan \& Pringle, 1998) and thus unacceptably
short. However KLOP showed that with $J_d < 2J_h$ the black
hole spin can counteralign with the disc, so that accretion
then occurs in a fully retrograde fashion. A randomly--oriented
sequence of such accretion episodes can thus keep the spin low, rather
than all aligning the hole very quickly and causing systematic spinup.

The condition $J_d \la J_h$ requires
\begin{equation}
M_d \la Ma\biggl({R_g\over R_d}\biggr)^{1/2},
\label{discmass}
\end{equation}
where $R_d$ is a typical outer disc radius, and $R_g = GM/c^2$.  This
agrees with the estimate (eq. 20) in KLOP, which explicitly uses the
AGN disc properties computed by Collin--Souffrin \& Dumont (1990). 
One could now follow the procedure outlined at the end of the last
Section to follow the hole's evolution under a random series of
accretion events obeying eq (\ref{discmass}). However a
straightforward argument shows the likely outcome. To simplify the
analysis we assume that all the episodes have the same mass and
angular momentum. We take the disc angular momentum as either
precisely aligned or counteraligned with ${\bf J}_h$, which follows
from the result that the alignment/counteralignment timescale is
shorter than the disc accretion timescale. Then $M_d, R_d$ are the
same for all episodes. Since spindown is more efficient than spinup, a
random sequence of events must tend to decrease $|a|$ for the hole
towards zero. However this process stops once we reach equality in eq
(\ref{discmass}), when further episodes cause $a$ to oscillate between
a small positive value
\begin{equation}
a \approx {M_d\over M}\biggl({R_d\over R_g}\biggr)^{1/2}
\end{equation}
and a negative value which is still smaller (i.e. closer to zero:
recall that spindown is more efficient than spinup). Clearly for
suitable choices of $M_d, J_d$ one can arrange to keep $|a|$ small
enough ($\la 0.5$) to ensure the low accretion efficiency required for
rapid mass growth, assuming an adequate supply of mass, i.e.
sufficiently frequent small-scale mergers.

For typical parameters if $J_d \sim J_h$ we find that $R_d \sim
8000R_ga^{10/19}$ (cf. KLOP, eq 20) and then the mass $M_d$ in each
event is of order 1\% of the hole's mass or less. Thus we require
$\sim 100$ such episodes to occur during the growth time $\sim t_{\rm
Edd}$.  Using the results of Section 2, we find that if each accretion
episode adds as much as $\Delta M = 0.18M$ to the hole, $a$ would
oscillate between values $\pm 0.3$, with accretion efficiencies
$\epsilon$ ranging from 0.049 to 0.069. With the smaller accretion
episodes required to satisfy $J_d \la J_h$ we find that with
e.g. $\Delta M = 0.016M$ the hole spin oscillates between $a = \pm
0.03$ with $\epsilon$ in the range $0.056 - 0.058$. This allows
a mass $M \sim 5\times 10^9\msun$ to grow in about $5\times
10^8$~yr, about one half of the time available at redshift $z =
6$. Growth of such a mass in a significantly shorter time would
require systematically retrograde accretion, which appears implausible.

\section{Conclusions}

We have shown that SMBH can grow their mass rapidly if most of it
comes from a sequence of randomly oriented accretion episodes whose
angular momenta $J_d$ are no larger than the hole's angular momentum
$J_h$. This comes about because the hole then has a low spin and thus
a low radiative accretion efficiency, so rapid mass growth is possible
without breaching the Eddington limit. The hole ends up with total
angular momentum comparable to that accreted during the last accretion
episode. Volonteri et al. (2005) suggested that most SMBH mass growth
occurs through a series of major galaxy mergers in each of which the
black hole accretes 2 -- 4 times its original mass. However,
cosmological simulations, including the ones upon which these
estimates were based, are not capable of following the detailed
hydrodynamics of the accretion process.  To satisfy the conditions
discussed here we need the accretion hydrodynamics to be chaotic at
the level of the size of the disc, that is on a scale of around $\la
0.1$ pc. Given that any such merger is likely to involve major
episodes of star formation, especially in the central regions, this
seems quite likely to be the case.

\section{Acknowledgments} 

ARK acknowledges a Royal Society--Wolfson Research Merit Award.

\label{lastpage}


\begin{thebibliography}{}

\bibitem{} Bardeen, J.M., 1970, Nat, 226, 64

\bibitem{} Bardeen, J.M., Petterson, J.A., 1975, ApJ, 195, L65

\bibitem{} Barth, A.J., Martini, P., Nelson, C.H., Ho, L.C., 2003, ApJ, 594,
L95

\bibitem{} Begelman, M.C., Volonteri, M., Rees, M.J., 2006, MNRAS, 370, 289
                                                                               
\bibitem{} Collin-Souffrin, S., Dumont, A.M., 1990, A\&A, 229, 292

\bibitem{} Hughes, S.A., Blandford, R.D., 2003, ApJ, 585, L101

\bibitem{} Kendall, P., Magorrian, J., Pringle, J.E., 2003, MNRAS, 346, 1078

\bibitem{} King, A.R., Lubow, S.H., Ogilvie, G.I., Pringle, J.E., 2005, 363, 49
                                                                   
\bibitem{} Lodato, G., Pringle, J.E., 2006, MNRAS, 368, 1196

\bibitem{} Madau, P., 2004, astro-ph/0410526.

\bibitem{} Moderski, R., Sikora, M., 1996, MNRAS, 283, 854

\bibitem{} Natarajan, P., Pringle, J.E., 1998, ApJ, 506, L97

\bibitem{} Scheuer, P.A.G., Feiler, R., 1996, MNRAS 282, 291 (SF96)

\bibitem{} Schmitt, H.R., Donley, J.L., Antonucci, R.R.J., Hutchings,
J.B., Kinney, A.L., Pringle, J.E., 2003, ApJ, 597, 768

\bibitem{} Soltan, A., 1982, MNRAS, 200, 115

\bibitem{} Volonteri, M., Madau, P., Quataert, E., Rees, M.J., 2005,
ApJ, 620, 69

\bibitem{} Volonteri, M., Rees, M.J., 2005, ApJ, 633, 624

\bibitem{} Willott, C.J., McLure, R.J., Jarvis, M.J., 2003, ApJ, 587, L15

\bibitem{} Wilson, A.S., Colbert, E.J.M, 1995, ApJ, 438, 62

\bibitem{} Yu, Q., Tremaine, S., 2002, MNRAS 335, 965

\end{thebibliography}
\end{document}